\documentclass[12pt,a4paper,dvips]{article}

\usepackage{amsmath}
\usepackage{bm}
\usepackage{graphicx}
\newcommand\TT{\rule{0pt}{2.5ex}}        
\newcommand\BB{\rule[-1.0ex]{0pt}{0pt}}  
\def\Xint#1{\mathchoice
{\XXint\displaystyle\textstyle{#1}}%
{\XXint\textstyle\scriptstyle{#1}}%
{\XXint\scriptstyle\scriptscriptstyle{#1}}%
{\XXint\scriptscriptstyle\scriptscriptstyle{#1}}%
\!\int}
\def\XXint#1#2#3{{\setbox0=\hbox{$#1{#2#3}{\int}$ }
\vcenter{\hbox{$#2#3$ }}\kern-0.6\wd0}}

\def\dashint{\Xint-}
\newcommand{\be}{\begin{equation}}
\newcommand{\en}{\end{equation}}
\newcommand{\bea}{\begin{eqnarray}}
\newcommand{\ena}{\end{eqnarray}}
\newcommand{\lbl}[1]{\label{eq:#1}}
\newcommand{\lbltab}[1]{\label{tab:#1}}

\newcommand{\lblsec}[1]{\label{sec:#1}}
\newcommand{\rf}[1]{(\ref{eq:#1})}
\newcommand{\Table}[1]{\ref{tab:#1}}
\newcommand{\fig}[1]{\ref{fig:#1}}
\newcommand{\sect}[1]{\ref{sec:#1}}
\newcommand{\braque}[1]{{\langle #1 \rangle}}
\newcommand{\bc}{\begin{center}}
\newcommand{\ec}{\end{center}}
\newcommand{\bt}{\begin{tabular}}
\newcommand{\et}{\end{tabular}}
\newcommand{\ba}{\begin{array}}
\newcommand{\ea}{\end{array}}
\newcommand{\gapprox}{%
\mathrel{%
\setbox0=\hbox{$>$}\raise0.3ex\copy0\kern-\wd0\lower0.80ex\hbox{$\sim$}}}

\newcommand{\lapprox}{%
\mathrel{%
\setbox0=\hbox{$<$}\raise0.6ex\copy0\kern-\wd0\lower0.65ex\hbox{$\sim$}}}
\newcommand{\inleft}{%
\mathrel{%
\setbox0=\hbox{$<$}\copy0\kern-0.5\wd0\lower1.1\ht0\hbox{$\scriptstyle{in}$}}}
\newcommand{\inright}{%
\mathrel{%
\setbox0=\hbox{$>$}\copy0\kern-0.5\wd0\lower1.1\ht0\hbox{$\scriptstyle{in}$}}}
\newcommand{\outleft}{%
\mathrel{%
\setbox0=\hbox{$<$}\copy0\kern-0.5\wd0\lower1.1\ht0\hbox{$\scriptstyle{out}$}}}
\newcommand{\outright}{%
\mathrel{%
\setbox0=\hbox{$>$}\copy0\kern-0.5\wd0\lower1.1\ht0\hbox{$\scriptstyle{out}$}}}

\newcommand{\im}{{\rm Im\,}}
\newcommand{\re}{{\rm Re\,}}
\newcommand{\Kbar}{\bar{K}}

\newcommand{\mpid}{m_\pi^2}

\newcommand{\mkd}{m_K^2}

\newcommand{\figwidth}{0.66666\textwidth}
\newcommand{\figwidthb}{0.53333\textwidth}

\newcommand{\MOM}{\bm{\Omega}}
\newcommand{\MOMinv}{\bm{\Omega}^{-1}}
\newcommand{\MT}{\bm{T}}

\begin{document}

\title{Couplings of  light $I=0$ scalar mesons to simple operators in
  the complex plane}

\author{B. Moussallam \\
{\sl Groupe de Physique Th\'eorique, IPN}\\
{\sl Universit\'e Paris-Sud 11, F-91406 Orsay, France} }


\maketitle

\begin{abstract}
The flavour and glue structure of the light
scalar mesons in QCD are probed by studying the couplings of the
$I=0$ mesons $\sigma(600)$ and $f_0(980)$ to the operators $\bar{q}q$,
$\alpha_s G^2$ and to two photons. 
The Roy  dispersive representation for the $\pi\pi$ amplitude
$t_0^0(s)$ is used to determine the pole positions as well as the
residues in the complex plane. 
On the real axis, $t_0^0$ is constrained to solve the Roy equation
together with elastic unitarity up to the $K\Kbar$ threshold leading
to an improved description of the $f_0(980)$. The
problem of using a two-particle threshold as a matching point is
discussed. 
A simple relation is established between the coupling of a scalar
meson to an operator $j_S$ and the value of the related pion
form-factor computed at the resonance pole. 
Pion scalar form-factors as well as two-photon partial-wave amplitudes
are expressed as coupled-channel Omn\`es dispersive
representations. Subtraction constants are constrained by chiral
symmetry and experimental data.
Comparison of our results for the $\bar{q}q$ couplings with earlier
determinations of the analogous couplings of the lightest $I=1$ and $I=1/2$
scalar mesons are compatible with an assignment of the $\sigma$, $\kappa$,
$a_0(980)$, $f_0(980)$ into a nonet.
Concerning the gluonic operator $\alpha_s G^2$ we find a significant
coupling to both the $\sigma$ and the $f_0(980)$.

\end{abstract}
\section{Introduction:}
Exotic hadrons in QCD remain  poorly understood theoretically. The
recent discoveries of the $X$, $Y$, $Z$ states~\cite{Choi:2003ue}, for
instance, in the charmonium spectroscopy was rather unexpected. Many
of the expected states, on the other hand, which are associated with
gluonic excitations like hybrids or  glueballs have not been
unambiguously identified.  The $0^{++}$ glueball is the lightest
stable particle in the QCD spectrum in the limit where all quark
masses are sent to infinity. In this situation, its mass has been
computed rather accurately in quenched lattice QCD
simulations~\cite{Morningstar} to be slightly smaller than two GeV. In
the presence of finite quark masses, the properties of the glueball
should remain relatively undisturbed provided $m_q \gapprox 1$ GeV. In
the physical situation, however, three quarks are substantially
lighter than 1 GeV.  Unquenched  lattice simulations have been
performed but the results are somewhat contradictory.
The simulations of ref.~\cite{Hart:2006ps} obtain near to maximal
mixing between glueball and $\bar{q}q$ states and find that
unquenching leads to a strong lowering of the masses. A similar effect
of unquenching was observed for  the $I=1$ scalar mesons by several
groups (e.g.~\cite{Frigori:2007wa,Hashimoto:2008xg}). This picture,
however, is not confirmed by the recent results from
ref.~\cite{Richards:2010ck} based on unquenched simulations with
$N_f=2+1$ and larger statistics who find gluball states  very similar
to the quenched ones.       

A possible scenario, suggested from using Laplace sum
rules~\cite{novikov,narisonveneziano} \footnote{references to more
  recent work which incorporate, in particular, more realistic
  modelling of instanton effects can be traced e.g. from
  ref.~\cite{Harnett:2008cw}} is that there could be two mesons below
2 GeV with  large glueball overlap. One of these could be rather light
and possibly identified with the $\sigma(600)$. Phenomenological
implications of this scenario have been discussed in some detail in
ref.~\cite{Minkowski}.   

The classification of the lowest lying experimentally observed scalar
mesons into a flavour nonet is also not a completely solved
problem~\cite{tornqvistrevue}. It has been proposed, for instance,
that the $a_0(980)$ and the $f_0(980)$ mesons could have a specific
status as weakly bound $K\Kbar$
molecules~\cite{Weinstein:1990gu}. This model simply explains their
near degeneracy and their proximity to the $K\Kbar$ threshold. It also
seems able to explain the values of the $2\gamma$ partial
widths~\cite{Hanhart:2007wa}.  Alternatively, it has been pointed out
a long time ago that the mass pattern of the nonet below 1 GeV can be
understood assuming a tetraquark flavour structure~\cite{Jaffe:1976ig}
(see also~\cite{Black:1998wt}).  

The peculiarity of a nonet composed of the $\sigma$, $\kappa$, $a_0(980)$
and $f_0(980)$ is most clearly formulated in terms of 't Hooft's large
$N_c$ limit of QCD~\cite{'tHooft:1973jz}. The masses, for instance, strongly
deviate from the ideal mixing pattern predicted in this
limit\footnote{In principle, dual ideal mixing is 
  possible~\cite{Black:1998wt}. The scalars must then be either
  tetraquarks, i.e. exotics, or else the mass squared of the
  $\sigma$-meson must be a decreasing function of the strange quark
  mass~\cite{Cirigliano:2003yq} which is unphysical.}. This implies
that in discussing the light scalars, effects which are sub-leading in
$1/N_c$, such as meson loops, ought to be taken into
account. Modellings  of meson loops effects can be found in the
classic papers~\cite{Tornqvist:1982yv,vanBeveren:1986ea}. More
recently, a model from which an explicit $1/N_c$ dependence can be
deduced has been proposed~\cite{Pelaez:2003dy}. Investigations in the
ADS/CFT modelling of large $N_c$ QCD have also been
performed~\cite{Colangelo:2008us}.  

Experiments on radiative decays of the $\phi$ meson have been
proposed~\cite{Achasov:1987ts} in order to clarify the flavour
structure of the light scalars. Such experiments have been performed
and are planned to continue (see~\cite{kloe2revue} for a review). The
simplest way, however, to quantify the various aspects of the
structure of the scalar resonances would be  via their couplings to a
set of simple operators. The glue content, for instance, is best
probed from the coupling to the gluonic operator $\alpha_s
G^2$. Similarly, the $\bar{q}q$ content is probed by the couplings of
the scalar mesons to quark-antiquark operators. Such couplings have
been considered for the $I=1$ and $I=1/2$ scalars by
Maltman~\cite{Maltman:1999jn} who suggested that their values can also
be used for properly identifying the nonet. A lattice QCD 
result for the coupling of $I=1$ scalars to $\bar{u}d$  is presented
in~\cite{McNeile:2006nv}. Studies of couplings to tetraquark operators
have  also been recently undertaken~\cite{Prelovsek:2010kg,Jansen:2009hr}.    
 
The $\sigma(600)$ resonance is very unstable and does not give rise to
a usual Breit-Wigner behaviour in cross-sections. Its existence has been
demonstrated only recently~\cite{CCL} by making a combined use of
experimental data and theoretical properties of the $\pi\pi$
scattering amplitude, which can be encoded into the set of
Roy~\cite{Roy:1971tc} integral equations. 
On the real axis, where the additional constraint of unitarity
applies, the Roy equations  were known as a powerful
tool for analyzing experimental pion-pion scattering 
data~\cite{pennington73,BFP74}. New high precision experimental  data
on low energy pion-pion
scattering~\cite{E865,NA48/2cusp,DIRAC,NA48/2Kl4,lastNA48} have
spurred renewed interest in these
equations~\cite{anantbuttiker,ACGL,DFGS,GKPY2,GKPY3}.   
In ref.~\cite{ACGL}, the
Roy equations are treated as a boundary value problem and exact
solutions have been searched for numerically below a matching point
$\sqrt{s_A}=0.8$ GeV. 

When applied to resonances, the Roy equations are used for computing the
partial-wave amplitude for complex values of the energy. The masses
and widths of the resonances
may be identified from the poles of the amplitude on the second Riemann sheet. 
The domain of validity of the Roy equations, as displayed in
ref.~\cite{CCL} allows one to discuss both the $\sigma$ and the
$f_0(980)$. The same poles which appear in the
elastic scattering amplitude can be shown to also appear in two-point
correlation functions of scalar operators and also in $\pi\pi$ matrix
elements of these operators. The poles  also appear in scattering
amplitudes with a pion pair in the final state like
$\gamma\gamma\to\pi\pi$. The residues of the poles are also determined
and can be interpreted in terms of couplings between scalar resonances and
operators. In the present work we consider, from this point of view,
the couplings of the scalar $I=0$ mesons $\sigma$ and $f_0(980)$ to the gluonic
operator $\alpha_s G^2$ and to the quark operators $\bar{u}u+\bar{d}d$
and $\bar{s}s$. We will update the results that can be obtained for
these couplings using the Roy equations combined with low-energy
constraints from chiral symmetry. We will also consider the couplings
to two photons, which were discussed in a similar framework in
ref.~\cite{Pennington:2006dg}. In that case, chiral constraints can be used
as well as recent experimental data from the Belle
collaboration~\cite{Belle1,Belle2}. 

The plan of the paper is as follows. We begin in sec.~\sect{royeq} by
constructing solutions to the Roy equations in a domain which extends
up to the $K\Kbar$ threshold. This domain covers most of the $f_0$
effect on the real axis.  We find that a very simple generalisation of
the parametrisations used in ref.~\cite{ACGL} is adequate for
approximating the solutions. In sec.~\sect{poles} we use these
solutions inside the Roy integral representations to perform
extrapolations to the complex energy plane. We determine the resonance
poles and their associated residues (sec.~\sect{poles}). These results are
applied in sec.~\sect{2gammas} to the determination of the scalar
mesons couplings to two photons. For this purpose, we use the
coupled-channel dispersive  Omn\`es representation for
$\gamma\gamma\to\pi\pi, K\Kbar$ and the chirally constrained fits
performed in~\cite{garciamartinmou}.  The couplings of the scalar
mesons to operators are finally considered in sec.~\sect{operators}. A
complex  plane definition is proposed from which a simple relation is
obtained between the couplings and pion scalar form-factors computed
at the resonance pole positions. Evaluations are made possible in this
case by using chiral constraints for the form-factors in combination
with coupled-channel Omn\`es representations~\cite{DGL90}.

\section{Roy equation solution for $t_0^0(s)$ up to the $K\Kbar$
  threshold}\lblsec{royeq}  
In order to improve the determination of the $f_0(980)$ properties, we begin
in this section by constraining  the $I=0$ $S$-wave amplitude $t_0^0(s)$ 
to satisfy the Roy equation up to the $K\Kbar$ threshold\footnote{We
neglect isospin breaking and take $m_K=(m_{K^+}+m_{K^0})/2$}. The Roy
equation reads
\bea\lbl{singleroy}
&& \re t_0^0(s)= a_0^0 + {s-4\mpid\over 12\mpid}(2a_0^0-5a_0^2)\\
&& + {1\over\pi}\dashint_{4\mpid}^{\infty} ds' \Big[ \im t_0^0(s')
\left( {1\over s'-s} + K_0(s',s)\right)\nonumber\\
&& + \im t_1^1(s') K_1(s',s) +\im t_0^2(s') K_2(s',s) \Big]
+d_0^0(s)\nonumber
\ena
where $a_0^0$, $a_0^2$ are the $I=0,2$ $S$-wave scattering
lengths. Detailed expressions for the kernels $K_a(s',s)$ and the
driving term $d_0^0(s)$ can be found in ref.~\cite{ACGL}.
Eq.~\rf{singleroy} is supplemented with the  non-linear,
unitarity relation involving the inelasticity
parameter $\eta_0^0(s)$
\be\lbl{unitrel}
\vert 1+ 2i\sigma_\pi(s) t_0^0(s) \vert=\eta_0^0(s)
\en
with $\sigma_\pi(s)=\sqrt{1-4\mpid/s}$. 
The inelasticity parameter $\eta_0^0$ is rigorously equal to one in
the region $s\le 16\mpid$.  Based on experimental indications, we use
here the approximation $\eta_0^0(s)=1$ up to the $s=4\mkd$. 
Furthermore in eq.~\rf{singleroy},  we use for the $P$-wave $\im(t_1^1)$
as well as the $I=2$ partial-wave $\im(t_0^2)$ inputs taken from
ref.~\cite{ACGL} i.e. satisfying the coupled Roy equations below 0.8
GeV and taken from experiment above. Imaginary parts of higher
partial-waves, which enter into the driving term $d_0^0$ are also
taken from experiment.    

\subsection{Multiplicity of the solutions:}
Taking the matching point as $s_m=s_K=4\mkd$, the Roy
equation~\rf{singleroy} admits a 
family of solutions~\cite{pomponiuwanders,atkinson,gasserwanders}
rather than a unique one.  
We will assume that the phase-shift at the $K\Kbar$ threshold satisfies
\be
\pi < \delta_K < {3\pi\over2}\ , \quad \delta_K\equiv \delta_0^0(s_K),  
\en
which implies~\cite{pomponiuwanders,atkinson,gasserwanders} a
two-parameter family of solutions\footnote{  
One assumes that the following set of inputs are given: the two
scattering lengths $a_0^0$, $a_0^2$, the phase-shift $\delta_0^0(s)$
above the matching point, the inelasticity function $\eta_0^0(s)$ and,
finally, the imaginary parts of the partial-waves $\im t_0^2(s)$ and
$\im t_{l\ge 1}^a(s)$.}.
In other terms, we must impose two
conditions in order to select a unique solution. As one condition, we
can fix the value of the phase-shift at one energy, for instance
the value of
\be\lbl{deltaAetK}
\delta_A\equiv \delta_0^0(s_A),\quad \sqrt{s_A}=0.8\ \hbox{GeV}. 
\en
In order to define a second condition, we consider the singularity
of the derivative of the phase-shift at the matching point. For a  generic
Roy solution, the divergence depends on the value of the 
phase-shift at the matching point in the following
way~\cite{pomponiuwanders,gasserwanders} 
\be\lbl{divergalpha}
\left.{d\over ds}\delta_0^0(s)\right\vert_{s\to s_m^-}\sim (s_m-s)^{\alpha-1},\quad
\alpha={2\delta_0^0(s_m)\over\pi}-2\ .
\en
In our case, the  matching point coincides with a two-particle
threshold, we expect the derivative of the phase-shift to exhibit a
square-root singularity
\be\lbl{divergundemi}
\left.{d\over ds}\delta_0^0(s)\right\vert_{s\to s_K^-}= A\,
(s_K-s)^{-{1\over2}}\ .
\en
This divergence is weaker than the generic matching point
divergence~\rf{divergalpha} provided the threshold phase-shift is not
too large,
\be\lbl{maxdeltak}
\delta_0^0(s_K) < 225^\circ\ .
\en
We will assume here that this condition is fulfilled. In this case, we
can use as a second condition that the phase-shift behaves as in
eq.~\rf{divergundemi} close to the $K\Kbar$ threshold. It is not
difficult to work out the explicit expression for the coefficient $A$
of the square-root singularity in eq.~\rf{divergundemi}.
For this purpose, let us consider the unitarity relation for  $\im t_0^0$ in
the region of the $K\Kbar$ threshold
\be\lbl{imt00}
\im t_0^0(s)= \sigma_{\pi}(s) \vert t_0^0(s) \vert^2
+ \theta(s-s_K)\sigma_K(s) \vert g_0^0(s)\vert^2\ 
\en
where $g_0^0(s)$ is the partial-wave $\pi\pi\to K\Kbar$ amplitude with
$I=0$, $J=0$. 
The principal value integration in the Roy equation~\rf{singleroy} generates
singularities associated with discontinuities of the derivative of
$\im t_0^0(s')$. Finite discontinuities lead to logarithmic divergences 
upon integration. The square-root divergence is generated from the
function $\theta(s'-s_K)\sigma_K(s')$. Performing the integration
analytically in the neighbourhood of the threshold one easily finds that
\be\lbl{A}
A=
  {\sigma_\pi(s_K)\vert g_0^0(s_K)\vert^2\over
  2\cos2\delta_K  \sqrt{s_K} } \ .
\en 

Once a solution is found for a given value of $\delta_A$, we can
compute the $\chi^2$ over the experimental data in the range
$[s_A,s_K]$ and then search for the value of $\delta_A$  which
minimises this $\chi^2$. In practice, the value of the phase-shift at
the $K\Kbar$ threshold, $\delta_K$ should be constrained by the data on
both sides of the matching point.
We can thus constrain both parameters $\delta_A$ and
$\delta_K$ by fitting the experimental data  using Roy equation solutions.

\subsection{Numerical approximations to the solution}
Let us denote by ${\cal R}[t_0^0]$ the right-hand side of
eq.~\rf{singleroy} and by $\epsilon(s)$ the difference between the
left and right-hand sides
\be\lbl{epsilon}
\epsilon(s)= {\cal R}[t_0^0](s)-\re t_0^0(s)\ .
\en
We construct numerical approximations to the phase-shift
in the range $4\mpid\le s\le 4\mkd$ using a simple modification of the
Schenk parametrisation~\cite{schenk91} compatible with 
eq.~\rf{divergalpha} 
\bea\lbl{paramsol}
&& \tan\delta_0^0(s)=\\
&& \quad \sigma_\pi(s) 
\left[ a_0^0 +\sum_1^N \alpha_i \left( {s\over s_\pi}-1\right)^i \right]
{s_\pi-s_0\over s-s_0}\,
{\sigma^K(s_\pi)+\beta\over \sigma^K(s)+ \beta}\nonumber
\ena
with $s_\pi=4\mpid$ and $\sigma^K(s)=\sqrt{s_K/s-1}$. 
This representation involves $N$  polynomial parameters $\alpha_i$ 
plus 2 parameters $s_0$ and $\beta$. The last factor generates a
square-root divergence in the derivative of $\delta_0^0$ as expected from
eq.~\rf{divergundemi}. In principle, the parameter $\beta$ could be
determined as a function of the known~\rf{A} coefficient $A$ in front
of the divergence. In 
practice, we have left it as a free parameter, adjusted such as
to help approximate the solution for $s$ close to $s_K$ but not necessarily
reproducing the exact limiting behaviour for $s=s_K$. We have checked that the
correct order of magnitude for $A$ is reproduced.

\begin{figure}[ht]
\bc
\includegraphics[width=\figwidth]{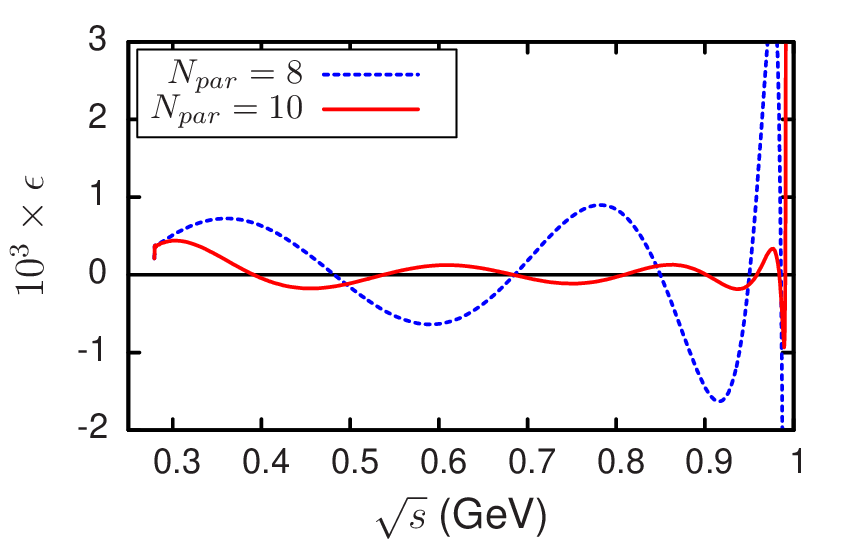}\\
\caption{\sl Error function (see eq.~\rf{epsilon}) corresponding to an
  approximation of the solution (see eq.~\rf{paramsol}) with 8
  parameters (dashed line) and 10 parameters (solid line)
}.
\label{fig:royerror}
\ec
\end{figure}

The $N+2$ parameters in eq.~\rf{paramsol} are determined from a
variational principle, by minimising the integral over the error
function squared  
\be
\chi^2_R\equiv \int_{4\mpid}^{4\mkd} ds' \left\vert
\epsilon(s')\right\vert^2 
\en
while fixing the two values of $\delta_0^0(s_A)$ and $\delta_0^0(s_K)$.
An exact solution corresponds to $\epsilon(s)$ vanishing identically 
in the whole range $[4\mpid,4\mkd]$ and therefore to $\chi^2_R=0$.
We used routines from the MINPACK library~\cite{minpack} to determine
the parameters in eq.~\rf{paramsol} which minimise $\chi^2_R$. 
We increased the number of parameters up to ten. With ten parameters 
one achieves an accuracy $\vert \epsilon(s)\vert\lapprox 5\, 10^{-4}$
below the matching point. The behaviour of the error function is illustrated in
fig.~\fig{royerror}. 
The figure shows that $\epsilon(s)$  is an oscillating
function which has a  number of
zeros approximately equal to the number of parameters in eq.~\rf{paramsol}.
The figure also illustrates  how the error function evolves
upon increasing the number of parameters which is suggestive of a
convergence towards an exact solution.
The accuracy is comparable to that quoted in ref.~\cite{ACGL}
below their matching point $s_A$.  Above the $K\Kbar$ threshold,
$\epsilon(s)$ increases rapidly becoming  $\simeq10^{-1}$. In this region,
this is similar to the results quoted in refs.~\cite{ACGL,GKPY2}. 
\vskip0.2cm 

\subsection{ Inputs above the matching point:}
The behaviour of the inelasticity function $\eta_0^0(s)$ close to the
$K\Kbar$ threshold is expected to have  a strong influence on the  
properties of the $f_0(980)$ resonance. 
By definition, $\eta_0^0(s)$
is equal to the modulus of the $\pi\pi\to\pi\pi$ partial-wave
$S$-matrix element. Unitarity of the $S$-matrix, 
\be
\vert S_{11}\vert^2=1-\sum_{n\ne1} \vert S_{1n}\vert^2 \
\en     
implies that $\eta_0^0\equiv \vert S_{11}\vert$ can be
determined experimentally either a) 
by measuring the cross-sections of the various open inelastic channels  
or b) by measuring the cross-section for elastic scattering.
The observation (by method (b)) that inelasticity sets in rather sharply at
the $K\Kbar$ threshold  suggests that  the 
$K\Kbar$ channel should  dominate the inelasticity  below
the $\eta\eta$ threshold. 
The $\pi\pi\to K\Kbar$ amplitude with $I=J=0$ has
been measured in high-statistics
experiments~\cite{cohen,etkin,Lindenbaum:1991tq}.  
We will use here the results of ref.~\cite{cohen} because the results
of~\cite{etkin,Lindenbaum:1991tq} have been argued to necessitate some
rescaling~\cite{morganpennington,buggzousarantsev}. The
$\pi\pi\to\eta\eta$ amplitude  has been measured in ref.~\cite{Alde:1985kp}.    
Some experimental information on the $\pi\pi\to 4\pi$ inelastic
amplitude is also available. We will rely on the discussion of
ref.~\cite{buggzousarantsev} who argue that the $\pi\pi\to 4\pi$
amplitude is small in magnitude below 1.4 GeV and can be modelled by 
contributions from the $f_0(1370)$ and the $f_0(1500)$ resonances.

\begin{figure}
\bc
\includegraphics[width=\figwidth]{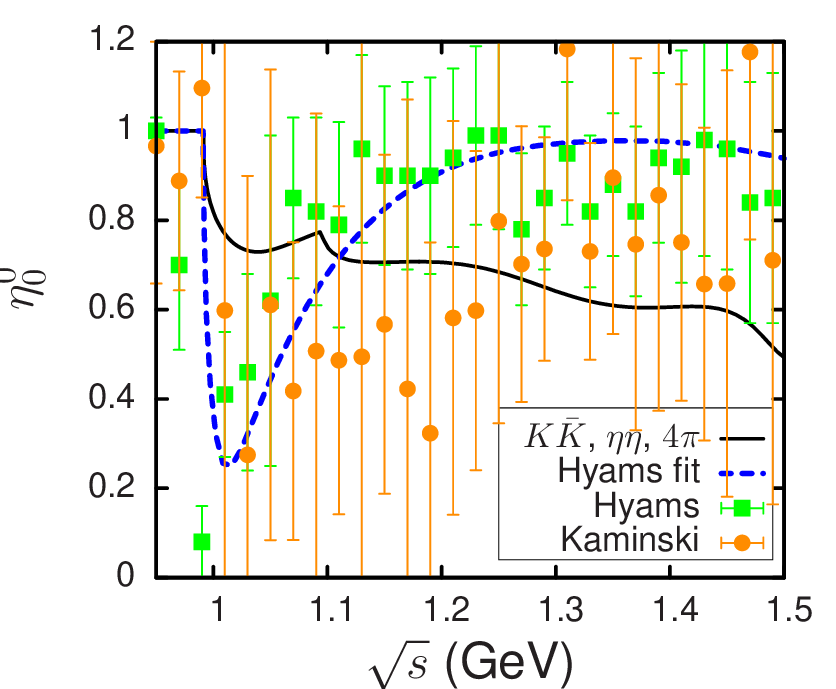}
\caption{\sl Inelasticity function $\eta_0^0(s)$. The data shown are
  determinations from the elastic amplitude
  from refs.~\cite{hyams73} and ~\cite{kaminski96}. The dashed
  line is the  $K$-matrix fit from ref.~\cite{hyams73}. The solid line is a
  determination of $\eta_0^0$ based on experimental information on
  the inelastic channels $K\Kbar$, $\eta\eta$ and $4\pi$. } 
\label{fig:inelastic}
\ec
\end{figure}
Fig.~\fig{inelastic} shows 
the experimental determinations of $\eta_0^0$ based on the elastic amplitude
from refs~\cite{hyams73} and ~\cite{kaminski96}.  The result of the $K$-matrix 
fit performed in ref.~\cite{hyams73} is plotted (dashed curve), which is
characterised by a rather deep dip near 1 GeV. 
We also show the central value of the fit based on the inelastic
channels (solid curve). The dip, in that case, is much less
pronounced. The inelastic determination of $\eta_0^0$ is actually not
inconsistent with the elastic determinations of
refs.~\cite{hyams73,kaminski96} within the errors.
It has a $\chi^2/N=1.6$  with the data of ref.~\cite{hyams73} and 
a $\chi^2/N=0.4$ with the data of ref.~\cite{kaminski96} (which is smaller
than one because of the very large errors).    
For the phase-shift $\delta_0^0$ above the $K\Kbar$ threshold,
we use the determination of Hyams et al.~\cite{hyams73}. It is in good
agreement with other analysis 
of the CERN-Munich experiment (e.g. ~\cite{buggzousarantsev}) 
or the analysis of the
CERN-Munich-Cracow experiment~\cite{kaminski96} below 1.5 GeV. The
energy region above 1.5 GeV is suppressed in the Roy equation because of the
two subtractions.

\subsection{Inputs below the matching point}
In the energy range $[s_A,4\mkd]$, in which we fit the two parameters
$\delta_A$ and $\delta_K$, we combine the sets of data 
from Hyams et al.~\cite{hyams73} and the data from Kaminski et
al.~\cite{kaminski96}. The former data have much smaller error
bars but it is likely that this is only because  Kaminski et
al.~\cite{kaminski96} have estimated their errors in a more realistic
way. This is suggested by comparing the phase-shifts resulting from different
analysis of the CERN-Munich experiment
(e.g.~\cite{estabrooks74,buggzousarantsev}, see also the
review~\cite{ochs91} for detailed comparisons and further
experimental references).  We have therefore 
appended a weight factor of $1/4$ to the $\chi^2$ of the data of Hyams
et al. in the combined $\chi^2$.    
For the $S$-wave scattering lengths, we take  the numbers quoted in the
latest NA48/2 publication~\cite{lastNA48}
\bea
&&a_0^0=0.2196\pm0.0028_{\hbox{stat}}\pm0.0020_{\hbox{syst}}\\
&&a_0^2=-0.0444\pm0.0007_{\hbox{stat}}\pm
0.0005_{\hbox{syst}}\pm0.0008_{\hbox{ChPT}}\nonumber 
\ena 

\begin{table}[htb]
\bc
\bt{c|cccc}\hline\hline
\TT\BB $\eta_0^0$ & $\delta_A$ & $\delta_K$ & 
$\hat{\chi}^2_{\hbox{\small\cite{hyams73}}}$ &
$\hat{\chi}^2_{\hbox{\small\cite{kaminski96}}}$\\ \hline
(a) & $\left(80.9\pm 1.4\right)^\circ$ & 
$\left(190^{+5}_{-10}\right)^\circ$ &2.7   &  1.9 \\  
(b) & $\left(82.9\pm 1.7\right)^\circ$   & 
$\left(200^{+5}_{-10}\right)^\circ$ &2.2   &  1.3\\ \hline\hline
\et
\caption{\sl Results for the two phases $\delta_A$ and $\delta_K$ from
  fitting the experimental phase-shifts in the range $0.8\ \hbox{GeV}
  \le \sqrt{s}\le 2m_K$ with Roy solution functions corresponding to
  two different central values of the inelaticity function (see
fig.~\fig{inelastic}). On the first line $\eta_0^0$ is determined from
a sum over inelastic channels (shallow-dip shape), on the second line
$\eta_0^0$ is determined from the elastic channel (deep-dip shape).
}  
\lbltab{fitres}
\ec
\end{table}

\begin{figure}[hbt]
\bc
\includegraphics[width=\figwidth]{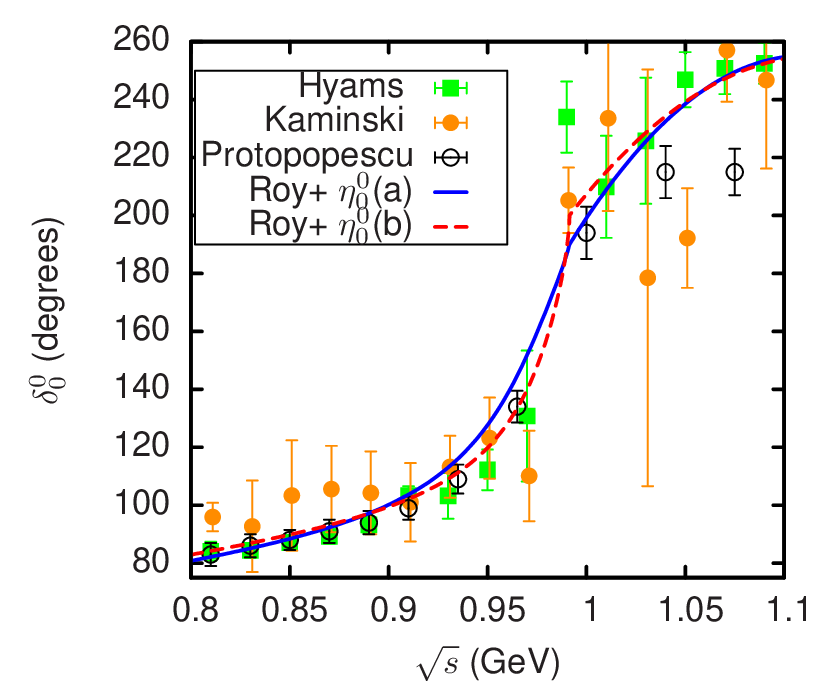}
\caption{\sl $I=0$ $S$-wave $\pi\pi$ phase-shifts: for $\sqrt{s}\le
  2m_K$ the two curves represent  solutions of the Roy equation
  corresponding to two different determinations of the inelasticity function
  $\eta_0^0$ (see fig.~\fig{inelastic}). 
}
\label{fig:comproyhy}
\ec
\end{figure}

The results of fitting the combined data sets as described above in the region
$[s_A,4\mkd]$ varying the two parameters $\delta_A$ and 
$\delta_K$ are presented in table~\Table{fitres}. 
We show separately
the result corresponding to the two different determinations of the
inelasticity function. We also show 
$\hat{\chi}^2=\chi^2/N$ (with $N=10$ data points) corresponding to
the data of Hyams\footnote{We remark that while the
$\chi^2$ seems large, half of its value comes from the single energy bin
with $E=0.99$ GeV. } et al.~\cite{hyams73}  
and to the data of Kaminski et al.~\cite{kaminski96}. The table shows
that a better $\chi^2$ is obtained upon using  the inelasticity
function from the elastic data (deep-dip shape). 
This reproduces the  observation first made in the recent
analysis of ref.~\cite{GKPY3}. In that work, a variant of the three
coupled Roy equations (derived from once-subtracted dispersion
relations) have been   considered  in their whole domain of validity,
i.e. up to $\sqrt{s}=1.1$ GeV and required to be satisfied
withing the errors of the data. Their analysis favours a value for the
threshold phase $\delta_K$ somewhat larger than the results of
table~\Table{fitres} while their result for $\delta_A$ is compatible
with ours. 
Fig.~\fig{comproyhy} displays the curves for the phase-shift
$\delta_0^0$ corresponding to the fit results of
table~\Table{fitres}. The figure also shows the phase-shifts from the
Berkeley experiment~\cite{protopopescu73} which were not included in
the fit. Numerical values of the parameters describing  the Roy
solution phase-shifts (see eq.~\rf{paramsol}) are given in the appendix.

\section{Poles and residues of the $\sigma(600)$ and
  $f_0(980)$}\lblsec{poles} 
Resonances correspond to poles of the $\pi\pi\to\pi\pi$ 
scattering amplitude $t_0^0(s)$ on unphysical Riemann sheets. These poles
are also present in form-factors and correlation functions which
involve currents which can couple to a pion pair in the $S$-wave.
We will consider only the second Riemann sheet here and recall a
few standard formulas which enable one to perform the
continuations~\footnote{More general formulas, for a four
sheets situation can be found e.g. in ref.~\cite{xiaozheng01}}.
These formulas can be expressed in terms of the amplitude $t_0^0(s)$. 

Let us start with the continuation of the amplitude $t_0^0$ itself. 
Its right-hand cut is associated with unitarity relations and has
successive thresholds in $s$: $4\mpid$, $16\mpid$, $36\mpid$,
$K\Kbar$, \ldots The second sheet is defined with respect to the
discontinuity relation which holds between the first two thresholds $4\mpid\le
s\le 16\mpid$.  Using the property of real-analyticity $t_0^0(z^*)=
{t_0^0}^*(z)$ (which results from $T$-invariance), it can be written as
\be\lbl{disct00}
t_0^0(s+i\epsilon)-t_0^0(s-i\epsilon)=2\sigma^\pi(s-i\epsilon)
t_0^0(s-i\epsilon)t_0^0(s+i\epsilon)
\en
where we have introduced 
\be
\sigma^\pi(z)\equiv \sqrt{4\mpid/z-1}\  
\en
(which satisfies $\sigma^\pi(s-i\epsilon)=i\sigma_\pi(s)$). From
relation~\rf{disct00}, one finds that the second sheet extension of
$t_0^0$ is 
\be\lbl{t00II}
t_0^{0,II}(z)= {t_0^0(z)\over 1- 2\sigma^\pi(z) t_0^0(z)}
\en
which is easily seen to verify the continuity relation
$t_0^{0,II}(s-i\epsilon)= t_0^0(s+i\epsilon)$.

The poles of the $T$-matrix can now be determined by searching for the
zeros of the denominator in eq.~\rf{t00II}: $S_0^0(z)= 1-
2\sigma^\pi(z) t_0^0(z)$ (which is the partial-wave $S$-matrix). The
derivative of $S_0^0(z)$ 
\be
\dot{S}_0^0(z_S)\equiv \left.{d\over dz} ( 1- 2\sigma^\pi(z)
t_0^0(z))\right\vert_{z=z_S} \ .
\en 
is needed in order to determine the residues. 
Numerical results for the pole positions and the $S$-matrix
derivatives are presented in table~\Table{zpoles}.
\begin{table}[htb]
\bc
\bt{c||cc}\hline\hline
\TT\BB & $\sqrt{z_S}$ (MeV) & $\dot{S}_0^0(z_S)$ (GeV$^{-2}$) \\ \hline
\TT\BB $\sigma(600)$ & $\left(442^{+5}_{-8}\right) +i\left(274^{+6}_{-5}\right)$ &  
$-\left(0.75^{+0.10}_{-0.15}\right)+i\left(2.20^{+0.14}_{-0.10}\right)$  \\
\TT\BB $f_0(980)$    & $\left(996^{+4}_{-14}\right) +i\left(24^{+11}_{-3}\right)$ & 
$-\left(1.1^{+3.0}_{-0.4}\right)+i\left(6.6^{+0.8}_{-1.0}\right)$   \\ \hline\hline
\et
\caption{\sl Positions of the complex poles and values of
the corresponding derivatives of the $S$-matrix $S_0^0$ from the Roy integral
representation of $t_0^0$ and the real-axis Roy solution discussed in
sec.~\sect{royeq}} 
\lbltab{zpoles}
\ec
\end{table}
The central values in the table  correspond to the Roy solution
associated with the deep-dip shaped $\eta_0^0$. This choice gives a
result for the $\sigma$ position very close to that of
ref.~\cite{CCL}. The errors were determined by varying the most
significant parameters in the Roy equation i.e. the two scattering
lengths $a_0^0$, $a_0^2$, the two phase-shifts $\delta_0^0(s_A)$,
$\delta_0^0(s_K)$ (see table~\Table{fitres}) and the parameters of the
$f_2$ meson which dominates the driving term. We have also included
the result of varying between the two different determinations of the
inelasticity in the form of asymmetric errors. For instance, using the
shallow-dip inelasticity, the value of the sigma pole position is
located at : $\sqrt{s_\sigma}=436+i278$ MeV and that of the $f_0$
is located at $\sqrt{s_{f_0}}=983+i36$ MeV.  The errors on the $\sigma$
pole parameters quoted in table~\Table{zpoles} are smaller than those
in ~\cite{CCL}: this can be traced to the fact than the range of
variation for the phase $\delta_0^0(s_A)$ as determined from the fit
using Roy solutions is smaller than the one estimated in ref.~\cite{CCL}. 

\section{Scalar meson couplings to two photons }\lblsec{2gammas} 

\subsection{$\gamma\gamma\to\pi\pi$ on the real axis:}
Informations on the couplings of the light scalar mesons to two
photons can be extracted from the amplitudes $\gamma\gamma\to
\pi^0\pi^0, \pi^+\pi^-$. This may be performed in a model independent
way by making use of the analyticity and unitarity properties of the
partial-wave amplitudes $h^I_{J,\lambda\lambda'}(s)$ which, as a consequence, satisfy
Omn\`es-type~\cite{omnes58} dispersive representations. A
representation of this kind for $h^0_{0,++}(s)$ was reconsidered
recently~\cite{garciamartinmou}  which makes use of a two-channel
extension of the Omn\`es approach~\cite{babelon76,zheng09}.
It should be valid in a range of energies up to one GeV where it is a
reasonably good approximation to retain just two channels ($\pi\pi$, 
$K\Kbar$) in the unitarity relation. This representation involves also the
$\gamma\gamma\to K\Kbar$ isoscalar partial-wave amplitude $k^0_{0,++}(s)$
and has the following form
\bea\lbl{2chanrepres}
&& \left(
\ba{l} 
h^0_{0,++}(s)\\
k^0_{0,++}(s)
\ea\right) =\\
&& \qquad\left(
\ba{l} 
\bar{h}^{0,Born}_{0,++}(s)\\
\bar{k}^{0,Born}_{0,++}(s)
\ea\right) + 
\MOM(s)\times\Bigg[
\left(\ba{l} 
b^{(0)} s +b^{'(0)} s^2\\
b_K^{(0)} s +b_K^{'(0)} s^2
\ea\right) 
\nonumber\\
&& \qquad
+{s^3\over\pi}\int_{-\infty}^{-s_0}{ds'\over(s')^3(s'-s)} 
\MOM^{-1}(s')\,\im
\left(\begin{array}{l} 
\bar{h}^{0,Res}_{0,++}(s')\\
\bar{k}^{0,Res}_{0,++}(s')
\end{array}\right)
\nonumber\\
&& \qquad
-{s^3\over\pi}
\int_{4\mpid}^\infty 
{ds'\over (s')^3(s'-s)} \im \MOM^{-1}(s')
\left(\begin{array}{l} 
\bar{h}^{0,Born}_{0,++}(s')\\
\bar{k}^{0,Born}_{0,++}(s')
\end{array}\right)
\Bigg]\ . \nonumber
\ena
The right-hand side of this equation involves 
the $2\times2$ Omn\`es matrix $\MOM$,
which encodes the effects of the final-state interaction. 
Its matrix elements $\MOM_{ij}$ are 
determined from the $T$-matrix by solving (numerically) the set of homogeneous
coupled integral equations  which arise from combining
dispersion relations and two-channel unitarity
\be\lbl{intomnes}
\Omega_{ij}(s)={1\over\pi}\int_{4\mpid}^\infty
{ds'\over s'-s} \left(\MT^*(s')\Sigma(s')
\MOM(s')\right)_{ij}
\en
with $\Sigma(s)=\hbox{diag}(\sigma_\pi(s),\sigma_K(s))$. One assumes
asymptotic conditions on $T_{ij}(s)$ (i.e. that
$T_{12}(s)$ goes to zero and that the sum of the
eigen-phase shifts goes to $2\pi$~\cite{mushkebook}) which ensure  that
eqs.~\rf{intomnes} have a unique solution once initial conditions are
specified  
\be
\Omega_{ij}(0)=\delta_{ij}\ .
\en
These asymptotic conditions are rather close from the experimental
values at $\sqrt{s}\simeq 2$ GeV.
Eq.~\rf{2chanrepres} also involves contributions from the
left-hand cut of the partial-waves which are associated
with singularities of the cross-channel amplitude $\gamma\pi\to
\gamma\pi$. The leading singularity arises from
the  charged pion pole which is exactly calculable and labelled 
$\bar{h}^{0,Born}_{0,++}(s')$ in eq.~\rf{2chanrepres}
(this term also dominates the amplitude in the soft photon limit). 
Singularities associated with multi-pion cuts are
described more phenomenologically (but with reasonable accuracy)
through the light resonance contributions, labelled
$\bar{h}^{0,Res}_{0,++}(s')$ in the above formula. Finally,
eq.~\rf{2chanrepres} involves four polynomial parameters. These have
been introduced by writing over-subtracted dispersion relations, such
as to cutoff integral contributions from higher energy regions.
The polynomial parameters have been determined in
ref.~\cite{garciamartinmou} from a chirally constrained fit\footnote{ 
The fit was performed in an energy range $\sqrt{s}\le 1.3$ GeV. For
$I=2$ amplitudes and for $J=2$ amplitudes, single channel Omn\`es
representations were used. Chiral constraints arise upon matching the
dispersive and the chiral two-loop
representations~\cite{gasserivan05,gasserivan06} from the fact that
the $p^4$ and certain $p^6$ chiral coupling-constants are known.}
of the experimental data from ref.~\cite{Belle1} (charged pions) and
ref.~\cite{Belle2} (neutral pions). 
In the present work, we use the $\pi\pi$ phase-shifts obtained by
solving the Roy equation below the $K\Kbar$ threshold in association
with the deep-dip inelasticity as discussed in sec.~\sect{royeq}. 
As compared to ref.~\cite{garciamartinmou}, this leads to small
differences in the $\gamma\gamma\to \pi\pi$ amplitudes localised in
the region of the $f_0(980)$ peak. The values of the fitted parameters
and the polarisabilities are not modified.    

\subsection{$\gamma\gamma\to\pi\pi$ in the complex plane}
\begin{figure}[ht]
\bc
\includegraphics[width=\figwidthb]{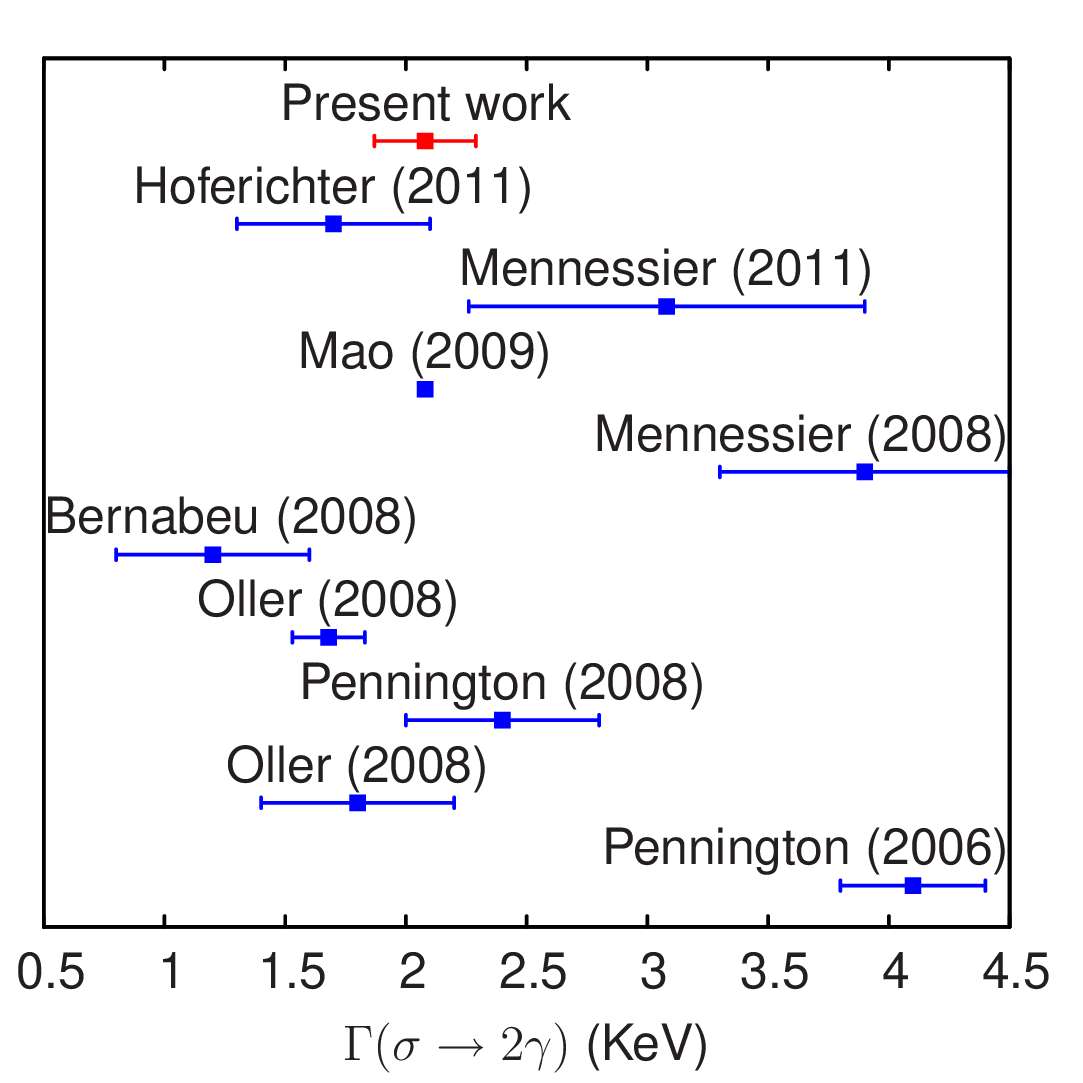}
\caption{\sl Recent determinations of the $\sigma\to 2\gamma$ width
  from experimental measurements of $\gamma\gamma\to 2\pi$ cross-sections.} 
\label{fig:sigma2g}
\ec
\end{figure}

Once the polynomial parameters are determined, the integral representations
~\rf{2chanrepres}~\rf{intomnes} allow one to compute the partial-wave
amplitude $h^0_{0,++}(s)$ for complex values of $s$. In order to
compute the second sheet extension one considers the discontinuity
between the first two thresholds which reads, 
\bea
&& h_{0,++}^0(s+i\epsilon)- h_{0,++}^0(s-i\epsilon)=\nonumber\\
&& \qquad 2\sigma^\pi(s-i\epsilon) t_0^0(s-i\epsilon) h_{0,++}^0(s+i\epsilon),
\ena
such that the second sheet extrapolation is 
\be\lbl{h00II}
h_{0,++}^{0,II}(z)= {h_{0,++}^{0}(z)\over 1- 2\sigma^\pi(z) t_0^0(z)}\ .
\en
The quantity of interest here is the decay width of the scalar mesons into
two photons. Following Pennington~\cite{pennington06}, it
can be defined by first identifying the residues of the amplitudes
$t_0^{0,II}(z)$ and $h_{0,++}^{0,II}(z)$ in terms of coupling constants
\be\lbl{residt00}
\left.32\pi t_0^{0,II}(z) \right\vert_{pole}= {g^2_{S\pi\pi}\over
  z_S-z},\quad  
\left.  h_{0,++}^{0,II}(z)  \right\vert_{pole}= {g_{S\pi\pi}
  g_{S\gamma\gamma} \over z_S-z}\ .
\en
The couplings $g_{S\gamma\gamma}$, $g_{S\pi\pi}$ are expected to be
complex numbers (see below).
One can formally define the decay width by taking the usual relation
between a coupling constant and the corresponding decay width
\be
\Gamma_{S\to 2\gamma}\equiv {\vert g_{S\gamma\gamma}\vert^2 \over 16\pi
  m_S}\ ,
\en
which yields the following numerical results for the two-photons
widths of the scalar mesons
\be\lbl{w2gamma}
\ba{lll}
\Gamma_{\sigma(600)\to 2\gamma}&=\left(2.08\pm 0.20\,
^{+0.07}_{-0.04}\right) \ &\hbox{(keV)}\\[1mm]
\Gamma_{f_0(980)\to 2\gamma}  &=  \left(0.29 \pm0.21 \,
^{+0.02}_{-0.07}\right) \ &\hbox{(keV)}\ .
\ea\en
The separation of the errors reflect the structure of the Omn\`es
representation~\rf{2chanrepres}. 
The first error is associated with varying the subtraction parameters
in eq.~\rf{2chanrepres}, i.e. it essentially reflects
the experimental errors in the two-photon cross-sections. The second
error is associated with the uncertainties in the Omn\`es matrix elements
coming from the $\pi\pi$ phase-shifts and inelasticities.
Fig.~\fig{sigma2g} compares our value for the sigma  width with
results quoted in the recent 
literature~\cite{pennington06,oller07,pennington08,bernabeu08,
mennessier08,mao09,Mennessier10,Hoferichter:2011wk}
(see also ~\cite{achasov08}) which are all based on the complex pole
definition. Evaluations using a Breit-Wigner definition can yield a
somewhat different result (e.g.~\cite{Fil'kov:1998np}).  In the case
of the $f_0(980)$, which is a rather narrow resonance, the two
definitions should give reasonably compatible results. The central value
which we find in~\rf{w2gamma} is practically identical to the one
quoted in the PDG~\cite{PDG}.   

Let us finally quote the corresponding  central complex values of the coupling
constants $g_{S\gamma\gamma}$, $g_{S\pi\pi}$ (in GeV),
\be\lbl{complexg}
\ba{ll}
g_{\sigma\gamma\gamma}=(-0.31+i0.60)\,10^{-2},\ & g_{\sigma\pi\pi}=1.12+i4.63\\
g_{f_0\gamma\gamma}=(\,\,\,0.38-i0.02)\,10^{-2},\ &    g_{f_0\pi\pi}=0.23+i2.79\ .
\ea
\en
It is striking that these couplings can be far from being real. It is
difficult to find a general physical interpretation for the phases of
the couplings but it is instructive to consider the case of  
a narrow resonance, i.e. when $\im z_S$ is small. 
In this situation, the Breit-Wigner approximation describes the
amplitude in the region of the zero and the corresponding pole of the
resonance   
\be\lbl{breitwigner}
S_0^0(z)\simeq S_B(k){ k-k_S\over k-k^*_S} 
\en
where $k$ is the $\pi\pi$ momentum, $k=\sqrt{z/4-\mpid}$, and $S_B$
is a slowly varying function. Neglecting contributions which are
quadratic in $\im z_S$, this representation gives the derivative at
$z=z_S$ as 
\be\lbl{dotSBW}
\dot S_0^0(z_S)\simeq {i\exp(2i\delta_0^0(\re z_S))\over 2\im z_S}
\en
Using this in the expressions for the residues, the coupling
$g_{S\pi\pi}$ gets expressed as
\be
{g_{S\pi\pi}^2\over 32\pi}\simeq {-\exp(-2i\delta_0^0(\re z_S))\im
  z_S\over\sigma_\pi(\re z_S)}\ , 
\en
i.e. the phase of $g_{S\pi\pi}$ is given in terms of the phase-shift
at the resonance mass 
\be\lbl{phasegpipi}
g_{S\pi\pi}=\vert g_{S\pi\pi}\vert \hbox{e}^{i\left({\pi\over2}
-\delta_0^0(\re z_S)\right)} 
\en 
and vanishes only in the absence of any non-resonant background
phase. In the case of the coupling $g_{S\gamma\gamma}$ one finds, at
leading order in $\im z_S$, 
\be
g_{S\pi\pi} g_{S\gamma\gamma}\simeq 2i\exp(-2i\delta_0^0(\re
z_S))h_0^0(\re z_S) \im z_S\ 
\en 
i.e. (using~\rf{phasegpipi}) 
\be\lbl{phasegpigaga}
\hbox{Phase}\,(g_{S\gamma\gamma})= \hbox{Phase}\,(h_0^0(\re z_S))-\delta_0^0(\re z_S) 
\en
which vanishes modulo $\pi$ when  $\re z_S$ is in the region of
applicability of Watson's theorem. These narrow width estimates for
the phases provide a reasonably good approximation for the $f_0(980)$
when its mass is located below the $K\bar{K}$ threshold (which is not
the case for our central value). 

\section{$\sigma(600)$ and $f_0(980)$ couplings to gluon and quark
  operators}\lblsec{operators} 
\subsection{Definitions}
The $\sigma(600)$ and $f_0(980)$ mesons have the same quantum numbers
$J^{PC}=0^{++}$ as the vacuum which are also those  expected for the lightest
glueball. One can characterise the gluon content of a scalar meson from
its coupling to the gluonic operator $\alpha_s G^2$. One may also
consider the trace of the energy-momentum tensor operator,
$\theta_\mu^\mu$, which is proportional to $\alpha_s G^2$ in the chiral limit. 
Correspondingly, two coupling constants $C_S^{GG}$, $C_S^{\theta}$
(with mass dimension) can be introduced
\be\lbl{gcontentdef}
\ba{ll}
\braque{0\vert {\alpha_s} G^{a\mu\nu} G^a_{\mu\nu}\vert S}& =m_S^2\, C_S^{GG} \\
\braque{0\vert \theta_\mu^\mu \vert S} & = m_S^2 \,C_S^{\theta}\\
\ea
\en 
where $S$ is either the $\sigma$ or the $f_0(980)$ meson. We will also
consider matrix elements associated with  scalar quark-antiquark
operators. It is convenient to use a  normalisation which remains well
defined in the chiral limit
\be\lbl{qcontentdef}
\braque{0\vert \bar{u}u+  \bar{d}d \vert S}  =\sqrt2 B_0 \,C_S^{uu},\quad
\braque{0\vert \bar{s}s \vert S} = B_0 \,C_S^{ss} \ .
\en
with $B_0=-\lim_{m_q\to0}\braque{0\vert\bar{q}q\vert0}/F_\pi^2$.  With this
convention, the couplings are renormalisation group invariant in the
chiral limit.

At first, it is necessary to clarify the meaning of such matrix
elements  since scalar mesons are resonances and not stable
one-particle states. One may use a complex plane definition,
which is rather natural here as it applies equally well to broad
resonances like the $\sigma$ or to ordinary narrow resonances. A
simple relation between the couplings $C^{j}_S$ and pion scalar
form-factors can be derived. For this purpose, let us  consider
two-point correlation functions  
\be
\Pi_{jj}(s)=i\int d^4x e^{ipx} \braque{0\vert T j_S(x) j_S(0)\vert 0}
\en
where $j_S(x)$ is one of the  scalar operators considered above.   
The correlator $\Pi_{jj}$ satisfies a K\"allen-Lehmann representation
(e.g. ~\cite{IZ}) 
\be
\Pi_{jj}(s)={s^3\over2\pi}\int_{4\mpid}^\infty ds' {\rho_{jj}(s')\over
 (s')^3( s'-s)}+ \alpha s^2 +\beta s+\gamma
\en
(written here with three subtractions) in which the spectral function is
given as a  sum over a complete set of states
\be\lbl{specsum}
(2\pi)^4\sum_n \delta^4(p_n-q)\vert \braque{0\vert j_S(0)\vert n}\vert^2
=\theta(q_0)\rho_{jj}(q^2)\ .
\en
The discontinuity of $\Pi_{jj}$ across the real axis in the range
$4\mpid\le s\le 16\mpid$ is generated by the two-pion states $n=\pi^a\pi^a$
in the sum~\rf{specsum} and it can be written as  
\be\lbl{discPijj}
\Pi_{jj}(s+i\epsilon)-\Pi_{jj}(s-i\epsilon)= {3\over 16\pi} 
\sigma^\pi(s-i\epsilon) F_j(s-i\epsilon) F_j(s+i\epsilon)\ .
\en
Here, $F_j$ is the form-factor associated with the two-pion   matrix
element of $j_S$ 
\be
\braque{0\vert j_S(0)\vert \pi^i(p)\pi^j(p')}=\delta^{ij}
F_j((p+p')^2)\ .
\en
In deriving eq.~\rf{discPijj} one makes use of the fact that 
$F_j(s)$ is itself a real-analytic function. It  
has a cut along the positive real axis, and its discontinuity in the
range $[4\mpid,16\mpid]$ reads
\be\lbl{discFj}
F_j(s+i\epsilon)-F_j(s-i\epsilon)=2\sigma^\pi(s-i\epsilon)
t_0^0(s-i\epsilon) F_j(s+i\epsilon). 
\en
From the discontinuity relations~\rf{discPijj}~\rf{discFj}, it is
simple to deduce the second sheet extensions of the form-factor
\be\lbl{FSII}
F_j^{II}(z) = {F_j(z)\over 1- 2\sigma^\pi(z) t_0^0(z)}
\en
and that of the  correlator $\Pi_{jj}$
\be\lbl{PiII}
\Pi^{II}_{jj}(z)= \Pi_{jj}(z)+ {3\over 16\pi} {\sigma^\pi(z) \left(
  F_j(z)\right)^2\over  1- 2\sigma^\pi(z) t_0^0(z)}\ .
\en
These expressions show that the form factor and the correlation
function on the second Riemann sheet have exactly the same poles $z_S$ as
the $T$-matrix. Considering the residue of the pole  provides a
natural identification for the resonance couplings $\braque{0\vert
  j_S\vert S}$,
\be
\left.\Pi_{jj}^{II}(z)\right\vert_{pole}\equiv {  \left(\braque{0\vert j_S\vert
    S}\right)^2\over z_S-z}\ ,
\en
which thus get expressed in terms of the $\pi\pi$ form-factor
evaluated at the position of the pole, 
\be\lbl{ffactorrel}
\braque{0\vert j_S\vert S}= 
\sqrt{{-3\sigma^\pi(z_S)\over16\pi\,\dot{S}_0^0(z_S)}} F_j(z_S)\ .
\en
One can verify that the interpretation of residues in terms of
coupling constants  satisfy consistency conditions. For instance, one
expects the residue of the  form-factor $F_j^{II}(z)$ to involve the
product of the two couplings $\braque{0\vert j_S\vert S}$ and
$g_{S\pi\pi}$ in the following way
\be\lbl{ffpole}
\left. F_j^{II}(z)\right\vert_{pole} = { 
\braque{0\vert j_S\vert S}\times g_{S\pi\pi}\over\sqrt3(z_S-z)}\ .
\en
It is easy to verify that this expression can be exactly recovered
using formulas~\rf {t00II},\rf{FSII},~\rf{PiII} for the second-sheet
extensions together with the definition of $g_{S\pi\pi}$ from the
residue of $t_0^{0,II}(z)$ and the definition of $\braque{0\vert j_S\vert
  S}$ from the residue of $\Pi_{jj}^{II}(z)$.  

In the limit of narrow resonances, one can express the couplings
$C_S^j$ in terms of the form-factor $F_j$ evaluated on the real
axis. For this purpose, one can write $F_j$ in the neighbourhood of
the resonance position as a function of the momentum $k$ 
\be
F_j(z)={\phi_j(k)\over k-k^*_S}
\en
displaying explicitly the pole on the second sheet. If the pole is
close to the real axis we can expand the function $\phi_j(k)$,
\be
\phi_j(k_S)=\phi_j(\re k_S)+i(\im k_S)\phi'(\re k_S)+\cdots
\en
which, to lowest order in $\im k_S$ leads to the approximation
\be\lbl{FjBW}
F_j(z_S)\simeq {1\over2} F_j(\re z_S)\ .
\en
Using also the expression for the derivative of the $S$-matrix in the
narrow width limit~\rf{dotSBW} one can express the couplings in terms
of quantities evaluated on the real axis
\be\lbl{BWCj2}
\left(\braque{0\vert j_S\vert S}\right)^2\simeq {3\over 16\pi}
\sigma_\pi(M^2_S)M_S\Gamma_S \,
\left(\hbox{e}^{-i\delta_0^0(M^2_S)} F_j(M^2_S)\right)^2
\ ,
\en
using $\re z_Z\simeq M^2_S$, $\im z_S= M_S\Gamma_S$. This expression
shows that the squares of the couplings $C_S^j$ must be real numbers
in the narrow width limit, provided $M_S$ is in the region of
applicability of Watson's theorem. The couplings themselves can be
either real or pure imaginary depending on whether the phase shift and
the phase of the form-factor are equal or differ by $\pi$. 

\subsection{Numerical results}
Analyticity and unitarity allows one to derive
Omn\`es representations for the form-factors, analogous to those for the 
$\gamma\gamma\to \pi\pi$ amplitude but much simpler because of the
absence of a left-hand cut. Let us briefly recall the derivation. Let
$\overline{F}(s)$ be a two-component vector formed from the pion and kaon
form-factors, 
\be
^t\,{\overline{F}(s)}=(F_j^\pi(s),{2\over\sqrt3}F_j^K(s))
\en
and multiply it with the inverse of the Omn\`es matrix\footnote{The
  determinant of the Omn\`es matrix can be expressed in analytical
  form: $\hbox{det}
 \MOM(s)=\exp\left({s\over\pi}\int_{4\mpid}^\infty 
ds' {\phi(s')\over s'(s'-s)}\right)$ with
$\phi(s')=\theta(4\mkd-s')\delta_0^0(s')+\theta(s'-4\mkd)\delta_{\pi\pi\to
  K\Kbar}(s')$ which shows that it does not vanish.}
\be
\overline{G}(s)\equiv\ \MOMinv(s) \overline{F}(s)\ .
\en
This multiplication removes part of the right-hand cut i.e. the
components of $\overline{G}(s)$ have a right-hand discontinuity which
vanishes in the range 
\be
\im \overline{G}(s) \simeq 0,\quad 4\mpid\le s \le s_2
\en 
where $s_2$ is the point above which two-channel unitarity is no
longer a good approximation.  By construction, the components of
$\MOM(s)$ behave as $1/s$ when $s\to\infty$ and a similar behaviour
is expected from the from-factors, such that $\overline{G}(s)$ should
satisfy a once-subtracted dispersion relation. In terms of
$\overline{F}$, it reads
\bea
&& \overline{F}(s)=\MOM(s)\Big[ 
\left(\ba{c}
\alpha\\
\beta\\
\ea\right) \nonumber\\
&&\quad + {s\over\pi}\int_{s_2}^\infty {ds' \over s' (s'-s) }\, \im\left(
  \MOMinv(s')\overline{F}(s') \right)\Big]\ .
\ena
In the range $s<<s_2$, the energy dependence of the integral may be
neglected and one ends up with the following representation for the
form-factors 
\be\lbl{omffactor}
\left(\ba{r}
F_j^\pi(s)\\[0.1cm]
{2\over\sqrt3} F_j^K(s)
\ea\right)= 
\left(\ba{cc}
\Omega_{11}(s) & \Omega_{12}(s)\\[0.1cm]
\Omega_{21}(s) & \Omega_{22}(s)
\ea\right)\left(\ba{c}
\alpha +\alpha' s\\
\beta  +\beta' s
\ea\right)\ .
\en
As the discussion above shows, it is valid for $s << s_2$.
Such representations were used and discussed in detail in ref.~\cite{DGL90}. 
In order to determine the polynomial parameters, one
can rely on chiral symmetry~\cite{DGL90}. As a first approximation, one
can use the chiral expansions of the form factors at order $p^2$ and
determine the polynomial coefficients by matching the $O(p^2)$ values
of $F_j^P(0)$, $\dot F_j^P(0)$   
\begin{table}[htb]
\bc
\bt{c||rc||rc}\hline\hline
\TT\BB $j_S$ & $F_j^\pi(0)$ & $\dot{F}_j^\pi(0)$ & $F_j^K(0)$ &
$\dot{F}_j^K(0)$ \\ \hline
\TT $m_u\bar{u}u+
m_d\bar{d}d$   &  $\mpid$ & $0$ & ${1\over2}\mpid$ & $0$ \\
\TT $m_s\bar{s}s$  &  $0$    & $0$ & $\mkd-{1\over2}\mpid$ & $0$ \\
\TT $\theta_\mu^\mu$ & $2\mpid$ & $1$ & $2\mkd$  & $1$ \\ \hline\hline
\et
\caption{\sl Pion and kaon form-factors associated with various
  operators $j_S$. The table shows their values at $s=0$  and the
  values of their derivatives at leading chiral order.}
\lbltab{chiralFj}
\ec
\end{table}
with those of the Omn\`es representation. These $O(p^2)$ values are
recalled in table~\Table{chiralFj}. 
The representation~\rf{omffactor} then allows one to compute the
form-factors for complex values of $s$ (with $\vert s\vert < s_2$) and
thus determine the values of the couplings between scalar operators and
scalar mesons from  residue relations like~\rf{ffactorrel}. 

The numerical values of the absolute values of couplings (the phases
will be shown later) of the $\sigma$ and $f_0(980)$ mesons to  the
$\bar{q}q$ operators  obtained in this manner are collected in
table~\Table{qqbarcoupl}.   
\begin{table}[htb]
\bc
\bt{c||cc}\hline\hline
\TT  & $\sigma(600)$  & $f_0(980)$   \\ \hline
\TT$\vert C^{uu}_S\vert$ (MeV)& $206\pm4 ^{+4}_{-6}$  & $82\pm31 ^{+12}_{-7}$ \\
\TT$\vert C^{ss}_S\vert$ (MeV)& $17\pm5  ^{+1}_{-7}$  & $146\pm44
^{+14}_{-7}$
 \\ \hline\hline
\et
\caption{\sl Absolute values (in MeV) of the  couplings of the
  $\sigma$ and $f_0(980)$ mesons to scalar $\bar{q}q$ 
  operators as defined in eq.~\rf{qcontentdef}. } 
\lbltab{qqbarcoupl}
\ec
\end{table}
In this table, the first error reflects the influence of higher order
chiral corrections  in the polynomial parameters. We have estimated
that the order of magnitude, relative to the $O(p^2)$ values, should
be $\simeq30\%$ for the corrections proportional to $m_s$, and
neglected the corrections proportional to $m_{u,d}$. As expected, a
larger uncertainty is generated for the $f_0(980)$ than for the $\sigma$.
The second error is associated with the uncertainties in the $\pi\pi$
and $K\Kbar$ $T$-matrix as reflected in the Omn\`es matrix elements.

A previous estimate of the $\bar{q}q$ coupling of the $\sigma$ meson,
using Breit-Wigner approximations, was given in
ref.~\cite{gardnermeissner} in the form 
$\braque{0\vert\bar{d}d\vert\sigma}=\sqrt{2/3}B_0/\chi$, with
$\chi=20$ GeV$^{-1}$, which is significantly smaller than our
result. Some results for the couplings of the $I=1$ and $I=1/2$ mesons
to $\bar{q}q$ operators can be found in the litterature. These
resonances are reasonably narrow, such that various definitions should
be equivalent and we can compare their values to those we found for
the $I=0$ mesons.     
We  normalise the couplings of these
mesons in accordance with eq.~\rf{qcontentdef} 
\be\lbl{coupl1}
\braque{0\vert\bar u s \vert K^*_0}= B_0 C^{us}_{K^*_0},\quad
\braque{0\vert\bar u d \vert a_0  }= B_0 C^{ud}_{a_0}\ .
\en  
An evaluation of the $a_0(980)$ coupling was performed in
ref.~\cite{Maltman:1999jn} using finite-energy sum rules (see also
ref.~\cite{Narison:1984jr}). Converted to the normalisation of
eq.~\rf{coupl1}, the result of~\cite{Maltman:1999jn}
reads,  
\be\lbl{Ca0}
\vert C_{a_0(980)}^{ud}\vert =197\pm37\ \hbox{MeV}
\en
which is remarkably similar to the coupling of the $\sigma$ meson
$C_\sigma^{uu}$ in table~\Table{qqbarcoupl}. 
The coupling $C_{a_0(980)}^{ud}$ is related to the coupling $c_m$
introduced in ref.~\cite{Ecker:1988te} by $C_{a_0(980)}^{ud}=4c_m$ and
can be estimated from its relation to the low-energy chiral coupling
constants~\cite{Ecker:1988te}, eventually supplemented with large
$N_c$ or chiral sum rule
constraints~\cite{JOP2,rosellsanzcillero}. These approaches yield
values in the range $C_{a_0(980)}^{ud}= [120,200]$ MeV. An unquenched
lattice QCD calculation has also been performed~\cite{McNeile:2006nv}
which gives: $C_{a_0(980)}^{ud}= [304,340]$ MeV. These values should
not be compared too litteraly to the preceeding ones because they
correspond to unphysical pion masses $m_\pi/m_\pi^{phys} \gapprox 5$
and only two dynamical flavours.   
An estimate for the $\kappa$ meson coupling $C^{us}_\kappa$ can 
be made following a similar approach  to that used here for the
$\sigma$ meson. One can compute the position of the complex pole and
the corresponding value of the $S$-matrix derivative from the
Roy-Steiner equations~\cite{descotesmou}. The central values which one
obtains in this way are 
\be
\sqrt{z_S}\simeq (658+i\,277)\ \hbox{MeV},\quad
\dot{S}_0^{1\over2}(z_S)\simeq (0.59+i\,2.03)\ \hbox{GeV}^{-2}
\en
The coupling can then be defined in terms of the $K\pi$ scalar
form-factor evaluated at $z_S$ (see ref.~\cite{ElBennich:2009da},
appendix C) and this gives
\be
\vert C_{\kappa(800)}^{us}\vert \simeq 156 \ \hbox{MeV}\ .
\en
Comparing now the couplings of the $I=0$
mesons from table~\Table{qqbarcoupl} to those of the $I=1,\ 1/2$ mesons
one observes that the values of  $C^{uu}_\sigma$,
$C^{us}_\kappa$, $C^{ss}_{f_0(980)}$, $C^{ud}_{a_0(980)}$ are rather
similar, the relative differences do not exceed $\simeq 20\%$. This is
compatible with an assignment of the mesons $\sigma$, $\kappa$,
$f_0(980)$, $a_0(980)$ into a nonet. Results on the couplings of the
heavier scalar mesons $a_0(1450)$ and $K^*_0(1430)$ are also
available. Ref.~\cite{Maltman:1999jn} gives 
\be
\vert C^{ud}_{a_0(1450)}\vert= 284\pm54\ \hbox{MeV},\quad
\vert C^{us}_{K^*_0(1430)}\vert= 370\pm20 \ \hbox{MeV} \ .
\en
The result for the $a_0(1450)$ was obtained from a finite-energy sum
rule and the one for the $K^*_0(1430)$ from a one-channel Omn\`es
representation. An evaluation using a two-channel representation and
complex pole definition was made in ref.~\cite{ElBennich:2009da} which
gives $\vert C^{us}_{K^*_0(1430)}\vert \simeq 282$ MeV. With the
normalisations used here, the couplings of the $a_0(1430)$ and
$K^*_0(1430)$ to quark-antiquark operators seem to be significantly
larger than those of the light scalars.     

\begin{figure}[htb]
\bc
\includegraphics[width=\figwidth]{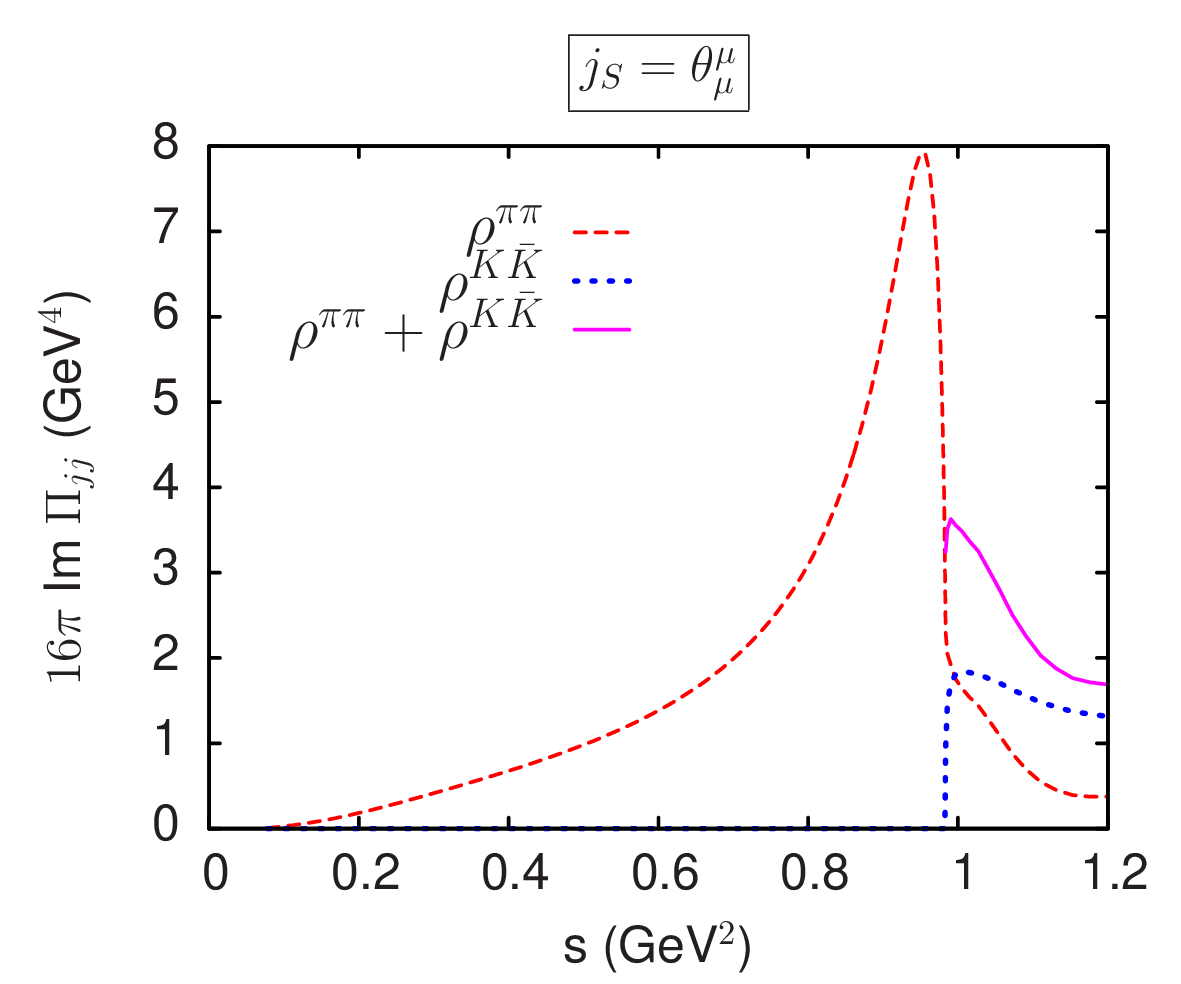}\\
\caption{\sl Spectral function of the $\Pi_{jj}$ correlator with
  $j_S=\theta_\mu^\mu$. The long-dashed and short-dashed curves are
  the contributions from the $\pi\pi$ and
  $K\Kbar$ intermediate states respectively. }
\label{fig:spectralmm}
\ec
\end{figure}

\begin{table}[htb]
\bc
\bt{c||cc}\hline\hline
\TT   & $\sigma(600)$  & $f_0(980)$   \\ \hline
\TT$\vert C^{\theta}_S\vert$ (MeV)& $197\pm15^{+21}_{-6}$& $114\pm44 ^{+22}_{-7}$ \\
\TT\BB$\vert C^{GG}_S\vert$ (MeV)& $472\pm 15 ^{+26}_{-16}$    & $227\pm41
^{+51}_{-16}$ \\ \hline\hline 
\et
\caption{\sl Absolute values of the couplings of the $\sigma$ and
  $f_0(980)$ to the gluonic operators $\theta_\mu^\mu$ and $\alpha_s G^2$.
} 
\lbltab{gluoncoupl}
\ec
\end{table}
Finally, one can compute the couplings of the light $I=0$ scalars to the
energy-momentum trace operator $\theta_\mu^\mu$ using the chiral
results for the associated form-factor at $s=0$ from
table~\Table{chiralFj} . The results are shown on the first line of
table~\Table{gluoncoupl}.   
One finds that both the $\sigma$ and the $f_0(980)$ display a
significant coupling to the $\theta_\mu^\mu$ operator. The trace of
the energy-momentum tensor has the following exact expression in
QCD~\cite{collins77}  with three heavy flavours integrated out
\be
\theta_\mu^\mu= {\beta(g)\over 2g} G^a_{\mu\nu} G^{a\mu\nu}
+(1+\gamma_m(g))\sum_{q=u,d,s} m_q \bar{q}q\ .
\en
This expression allows one to disentangle the $\alpha_s G^2$ part from the
$\bar{q}q$ one if one uses a perturbative approximation for the
$\beta$ function and for the anomalous dimension. The results shown in
table~\Table{gluoncoupl} for $C_S^{GG}$ correspond to a leading order
approximation. 
Our results for $C_S^\theta$  may be compared with the Laplace sum
rule evaluation~\cite{narisonveneziano} 
\be\lbl{narisonvenez}
C_\sigma^\theta=[272,329]\ \hbox{MeV}\ .
\en  
However, one should keep in mind that in the
calculation of~\cite{narisonveneziano}, the spectral
function $\im\Pi_{jj}(s)$ corresponding to the operator
$j_S=\theta_\mu^\mu$ is approximated by a simple delta
function. Fig.~\fig{spectralmm} shows our result for this spectral
function based on using two-channel unitarity and physical $\pi\pi$
scattering inputs. It displays a peak corresponding to the
$f_0(980)$ resonance, while the $\sigma$ resonance does not show up as
a clear enhancement, but  generates a broadening of  the
$f_0(980)$ peak at low energies. It is then plausible that the
value~\rf{narisonvenez} should be compared with the sum
$C_\sigma^\theta+C_{f_0}^\theta$ from table~\Table{gluoncoupl}: the
agreement is then rather reasonable.

Finally, the central values of the phases of the couplings $C_S^j$ are
shown in table~\Table{phasesCj}. In the Breit-Wigner approximation,
one expects the phases to be either zero of $\pm90^\circ$
(see~\rf{BWCj2}). The actual values are often not too different from
this approximation. 
\begin{table}[htb]
\bc
\bt{c|rr}\hline\hline
\TT                 & $\sigma\phantom{(6)}$ & $f_0(980)$ \\ \hline
\TT $\bar{u}u+\bar{d}d$ & $28.2^\circ$  & $ 89.1^\circ$ \\
\TT $\bar{s}s         $ & $-80.2^\circ$ & $-14.2^\circ$ \\
\TT $\theta_\mu^\mu$      & $87.2^\circ$  & $-34.6^\circ$ \\ \hline\hline 
\et
\caption{\sl Central values of the phases of the couplings $C_S^j$.}
\lbltab{phasesCj}
\ec
\end{table}
\section{Conclusions}
We have considered several properties of the light scalar isoscalar
mesons $\sigma$ and $f_0(980)$ using definitions which rely on the
positions of the poles in the complex plane and their associated
residues. This approach allows one to deal with a broad resonance like
the $\sigma$ in a well defined way. 
In order to compute the positions of the poles and  the residues,the
Roy integral representation for the $\pi\pi$ scattering amplitude
$t_0^0$ was used. On the real axis, we have started from the Roy
equation solutions of ref.~\cite{ACGL}, which use a matching point
$\sqrt{s_m}=0.8$ GeV and construct an extended solution which, for the
$S$-wave $t_0^0$, has a higher matching point  $\sqrt{s_m}=2m_K$ such
as to improve the theoretical constraints on the $f_0(980)$ meson
properties. In order to constrain the value of the $S$-wave scattering
phase-shift at the $K\Kbar$ threshold and discriminate between
different shapes of the inelasticity, corresponding to different
experiments, we perform fits of the phase-shifts below the $K\Kbar$
threshold based on the Roy solutions. We find that the solution
corresponding to a deep-dip shaped inelasticity has a better $\chi^2$
than that corresponding to a shallow-dip shape. This is in agreement with
the observations of ref.~\cite{GKPY3}. The properties of the $f_0(980)$
resonance, as expected, are particularly sensitive to the central
value of the inelasticity. The results based on this Roy representation
of the amplitude for the second-sheet pole positions are in
table~\Table{zpoles}.  

As a first application, we have re-determined the scalar to two photons
couplings $g_{S\gamma\gamma}$, following the methodology first
advocated in ref.~\cite{Pennington:2006dg}, and based on the
determinations of the $\gamma\gamma\to\pi\pi$ amplitudes from the
recent experimental measurements~\cite{Belle1,Belle2}. The result
found for the $\sigma$ is somewhat smaller than that originally given in
ref.~\cite{Pennington:2006dg}. 
As a second application, the couplings of the  $\sigma$ and $f_0(980)$
mesons to scalar operators, which can be formally denoted as
$\braque{0\vert j_S(0)\vert \sigma}$, $\braque{0\vert j_S(0)\vert
  f_0(980)}$ were defined and evaluated. 
Choosing $j_S=(\bar{u}u+\bar{d}d)/\sqrt2$, $j_S=\bar{s}s$ these
matrix elements provide a quantitative measure of the quark-antiquark
contents of the scalar mesons, while choosing $j_s=\theta_\mu^\mu$ is a
measure of the glue content. A simple, general relation can be
established between such couplings and the value of the pion
form-factor associated with the operator $j_S$ computed at the
position of the resonance  pole, $F_j^{\pi\pi}(z_S)$. This relation is
given in eq.~\rf{ffactorrel} in the general case and in eq.~\rf{BWCj2}
in the limiting case of a narrow resonance. 
Such form-factors are known to be calculable from a coupled-channel
Omn\`es representation~\cite{DGL90} which should be valid in a
complex energy range which accomodates the $\sigma$ as well as the
$f_0(980)$ resonances. The polynomial parameters in such
representations are constrained by chiral symmetry, for both the
$\bar{q}q$ and $\theta_\mu^\mu$ operators, and can be estimated from
the leading order chiral Lagrangian~\cite{DGL90}. In principle,
matrix elements of other types of operators, for instance tetraquark
operators, could be addressed in the same way. The values of
$F_j(0)$ and $\dot{F}_j(0)$, in such cases, are not predicted from
chiral symmetry but could be obtained e.g. from lattice QCD.   

The numerical results for the $\bar{q}q$ coupling constants of
$\sigma$ and the $f_0(980)$ mesons are shown in
table~\Table{qqbarcoupl}. The couplings are not particularly
suppressed but it would be interesting to compare them with couplings
to tetraquark operators. The couplings can also be compared  to the
analogous couplings of the $I=1$ and $I=1/2$ mesons to the $\bar{u}d$
and $\bar{u}s$ operators respectively for which estimates can be found
in the litterature including one calculation in lattice
QCD~\cite{McNeile:2006nv}. This comparison supports a nonet assignment
of the $\sigma$, $\kappa$, $a_0(980)$, $f_0(980)$ mesons. Our results
for the couplings to the gluonic operators $\theta_\mu^\mu$ and
$\alpha_s G^2$ indicate that both the $\sigma$ and $f_0(980)$ couple
significantly  to such operators as well.

\section*{Acknowledgements}
I would like to thank prof. W. Ochs for sending me original data tables  
and Martin Hoferichter for making several very useful comments on the
manuscript. 
\section*{Appendix}
We show below central values of the parameters of Roy solutions for
the phase-shift $\delta_0^0(s)$, for a ten-parameter approximation
according to eq.~\rf{paramsol}, corresponding to two different central
values of the inelasticity function $\eta_0^0$, see sec.~\sect{royeq}. 
\bc
\bt{lll}\hline\hline
\TT\BB             &  $\eta_0^0$: deep-dip   & $\eta_0^0$: shallow-dip \\ \hline
\TT $s_0$          & $0.724237452  $ & $0.736126142  $ \\
$\beta$        & $0.104114178  $     & $0.360170063  $ \\
$\alpha_1$     & $0.140785825  $ & $ 0.146890648 $ \\
$\alpha_2$     & $-0.0408980664  $ & $-0.0391129286  $ \\
$\alpha_3$     & $0.00648917902  $ & $ 0.00545496306  $ \\
$\alpha_4$     & $-0.000845352717  $ & $-0.000636406542  $ \\
$\alpha_5$     & $7.20101833 10^{-5} $ & $4.62727765  10^{-5}$ \\
$\alpha_6$     & $-2.89568524 10^{-6} $ & $-8.84679012 10^{-7}$ \\
$\alpha_7$     & $-8.92462472 10^{-9} $ & $-9.93513196 10^{-8}$ \\
$\alpha_8$     & $3.07108997 10^{-9} $ & $4.83952846 10^{-9} $ \\ \hline\hline
\et
\ec

\end{document}